# An hierarchical artificial neural network system for the classification of transmembrane proteins

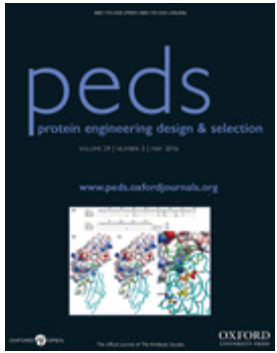

**Pasquier, C. and Hamodrakas, S.J.**

# An hierarchical artificial neural network system for the classification of transmembrane proteins

**Pasquier, C. and Hamodrakas, S.J.**

*Faculty of Biology, Department of Cell Biology and Biophysics,
University of Athens, Panepistimiopolis, Athens 15701, Greece*

**Abstract**

This work presents a simple artificial neural network which classifies, from their sequences alone, proteins into classes: the membrane protein class and the non-membrane protein class. This is important in the functional assignment and analysis of Open Reading Frames (ORF) identified in complete genomes and, especially those ORF's that correspond to proteins with unknown function.

The network described here have a simple hierarchical feed-forward topology and a limited number of neurons which make it very fast. By using only information contained in 11 protein sequences, the method was able to identify with 100% accuracy all membrane proteins with reliable topologies collected from several papers in the literature. Applied to a test set of 995 soluble proteins, the neural network classifies falsely 23 of them in the membrane protein class (with considerable success of 97.7% of correct assignment). The method was also applied to the whole SWISS-PROT database and on ORF's of several complete genomes.

The neural network developed was associated with the PRED-TMR algorithm (Pasquier et al., 1999) in a new application package called PRED-TMR2. A WWW server running the PRED-TMR2 software is available at http://o2.db.uoa.gr/PRED-TMR2

# Introduction

The number of protein sequences stored in public databases (78197 in SWISS-PROT release 37, 178773 in TrEMBL ; Bairoch and Apweiler, 1998) is considerably larger than that of known protein structures (9129 in PDB ; Sussman et al., 1998) and the gap is continuously increases as the experimental determination of the three-dimensional structure of proteins is a very time consuming process compared to the time needed for the determination of protein sequences. This is especially true for transmembrane proteins which are difficult to solve by X-ray crystallography.

Usually, the structure of a new protein having homologies above a certain level to another sequence of known structure can be predicted with reasonable accuracy (Persson and Argos, 1994; Rost et al., 1994; Rost et al. 1995). However, the majority doesn't belong to this ideal case. For this set of proteins, predictions methods that do not depend on sequence alignments but using solely information contained in a sequence itself is still needed.

A number of methods or algorithms designed to locate the transmembrane regions in proteins without the need for multiple-sequence alignment information have been developed (von Heijne, 1992; Cserzo et al., 1997; Pasquier et al., 1999). However, these algorithms focus on the localization of transmembrane segments in known integral membrane proteins and are not suited to be applied on soluble proteins.

Recently, we have published the PRED-TMR method in an attempt to improve the fine localization of transmembrane segments, by coupling an hydrophobicity analysis with a detection of potential termini (starts and ends) of transmembrane regions (Pasquier et al., 1999). Now, we have extended this application with a pre-processing stage represented by an artificial neural network which attempts to classify proteins to membrane and non-membrane.

Several applications of neural networks applied to the prediction of transmembrane segments or secondary structure prediction can be found in the literature (Reczko, 1993; Rost et al., 1994; Fariselli and Casadio, 1996; Aloy et al., 1997; Diederichs et al., 1998). Most of them use a local encoding for each amino acid and produce as output a classification for the amino acid in the middle of the input window. When a hierarchical feed-forward topology is used (the connectivity graph contains no loop), each network output is independent of the results obtained by previous processing. This causes no or little problem when the output of the network consists of continuous values (coordinates for example ; Diederichs et al., 1998). However, when a threshold parameter is used for the choice of binary output, the absence of correlation between the possible structure of adjacent residues frequently results in incoherent topologies (a transmembrane segment composed of only one residue for example). Authors solve this problem by designing recurrent neural networks which use additional information obtained with the processing of previous pattern (Reczko, 1993) or by building a system of cascading neural networks (Rost et al., 1994; Fariselli and Casadio, 1996). Nevertheless, these techniques are not appropriate for a correct classification between membrane and non-membrane proteins because they are too focused on one-residue topology prediction.

This paper presents an artificial neural network which doesn't predict the exact location of transmembrane segments but produces instead a unique output showing whether an analyzed part of a sequence is related to a transmembrane region or not.

# Methods

*Information gathering*

11 proteins, with known topologies were used for the training of the network : 6 transmembrane proteins containing a total of 19 transmembrane segments, 2 fibrous proteins and 3 globular ones. The sequences used and other information concerning the application are presented on our web site at http://o2.db.uoa.gr/PRED-TMR2/Results/).

Five different test data sets of transmembrane proteins with reliable topologies were collected from the literature. Test set 1 includes 64 sequences from the set of Rost et al. (1995) (the sequences 2MLT, GLRA_RAT, GPLB_HUMAN, IGGB_STRSP and PT2M_ECOLI which were not found in the public databases were not used), sets 2 and 3, of 48 and 83 proteins respectively were specified in Rost et al. (1996) paper, set 4 comprises the 44 sequences used by Cserzo et al. (1997) and set 5 is composed of 92 sequences from Fariselli and Casadio (1996).

A test data set of globular proteins was extracted from the Protein Data Bank (PDB), using the list of non homologous sequences of PDBSELECT (Hobohm et al., 1994). The 25% threshold list was used, excluding entries of membrane and lipid associated protein (1AIJ, 1ALY, 1AR1, 1ATY, 1BEH, 1BHA, 1BQU, 1BXM, 1FTS, 1IXH, 1JDW, 1KZU, 1LGH, 1LML, 1NKL, 1OCC, 1PRC, 1QCR, 1SQC, 1TLE, 1XDT, 1YST, 2CPS, 2MPR, 2OMF, 2POR AND 7AH1). This set of water-soluble proteins consists of 995 sequences.

**Table I.** Propensity values and corresponding input used in the neural network for the 20 amino acid residue types to belong to transmembrane segments, calculated from the entire SwissProt database.

| Residue | | $P_i$ | NN input |
|---|---|---|---|
| Phenylalanine | F | 2.235 | 1.000 |
| Isoleucine | I | 2.083 | 0.929 |
| Leucine | L | 1.845 | 0.817 |
| Tryptophan | W | 1.790 | 0.791 |
| Valine | V | 1.756 | 0.775 |
| Methionine | M | 1.502 | 0.655 |
| Alanine | A | 1.383 | 0.599 |
| Cysteine | C | 1.202 | 0.514 |
| Glycine | G | 1.158 | 0.494 |
| Tyrosine | Y | 1.075 | 0.455 |
| Threonine | T | 0.879 | 0.362 |
| Serine | S | 0.806 | 0.328 |
| Proline | P | 0.597 | 0.230 |
| Histidine | H | 0.395 | 0.135 |
| Asparagine | N | 0.389 | 0.132 |
| Glutamine | Q | 0.273 | 0.078 |
| Aspartic acid | D | 0.153 | 0.021 |
| Glutamic acid | E | 0.131 | 0.011 |
| Arginine | R | 0.124 | 0.007 |
| Lysine | K | 0.108 | 0.000 |

*Calculation of amino acid residue transmembrane propensities (potentials)*

As described by Pasquier et al. (1999), a propensity for each residue to be in a transmembrane region was calculated using the formula:

$$Pi = \frac{Fi^{TM}}{Fi}$$

where $Pi$ is the propensity value (transmembrane potential) of residue type $i$ and $Fi^{TM}$ and $Fi$ are the frequencies of the ith type of residue in transmembrane segments and in the entire SWISS-PROT database respectively. Values above 1 indicate a preference for a residue to be in the lipid-associated structure of a transmembrane protein, whereas propensities below 1 characterize unfavourable transmembrane residues. The propensity values for the 20 amino acid residue are given in Table I.

*Neural network topology and training parameters*

The neural network used here has a multi-layer feed forward (MLFF) topology. It consists of an input layer, one hidden layer and an output layer. Each of the units in the input layer are connected to all the units in the hidden layer. The units in the hidden layer are then connected to all the units in the output layer. This is a 'fully-connected' neural network where each unit $i$ of a given layer is connected to each unit $j$ of the next layer (Figure 1). The strength of each connection is given by a weight $w_{ij}$. The state $s$ of each unit in the input layer is assigned directly from the input data, whereas the states $s_j$ of higher layers $j$ are computed by the sigmoid function:

$$s_j = \frac{1}{1+e^{-(w_{j0}+\sum_{i=1}^{n} w_{ij}s_i)}}$$

where $w_{j0}$ is a bias from the states $s_i$ of lower layers.

**Fig. 1:** Schematical architecture of the neural network. Amino acids of the input sequence are converted to unique input values corresponding to the propensity for each amino acid to be located inside à transmembrane region. Output of the network consists of values between 0 and 1. Values above 0.9 (shown in black on the figure) indicate a detection of a potential transmembrane segment.

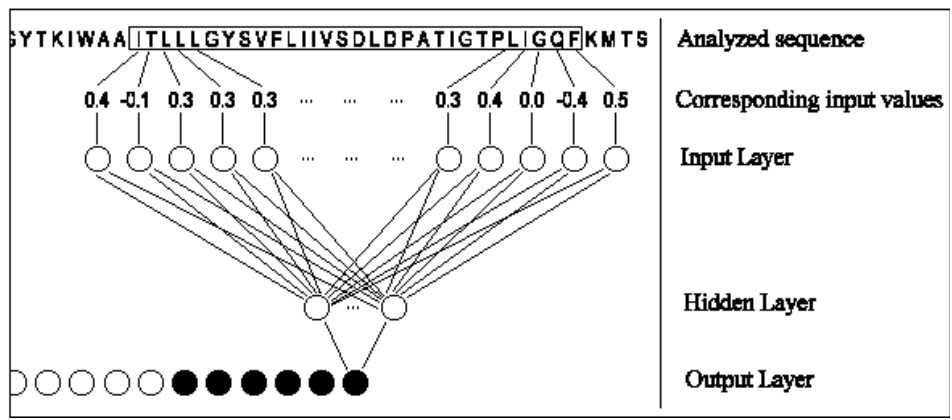

The network was trained using the backpropagation algorithm. During this process, a data set, describing the states $s_i$ of the input units and their desired output value is presented to the network. The activations of the units of the network are then calculated, feeding forward layer-by-layer from the inputs to the output. Once the network output value has been produced, it is compared to the target output specified in the training data set. Following this comparison, a backwards adjustment of the weights (backpropagation) is performed in order to minimize the differences between the computed output and the desired output value. The algorithm is performed until the total error reaches a low enough value which means that the network comes to approximate the target values given the inputs in the training set.

During the prediction phase, the neural network is fed with new input data that are not in the training set. By a simple feed-forward process, using the previously obtained weight, new output values are calculated and are taken as predictions of the network.

Applied to our classification problem, the idea is to use as input of the network a representation of a part or a sequence in order to obtain an unique output showing whether the analyzed segment is related to a transmembrane region or not.

The propensity values for the 20 amino acids given in Table I, which can be regarded as numeric representations of amino acids, are used to encode the input segment after being linearly transformed to lie in the range of 0 to +1. The output of the network consists of an unique value between 0 to 1 which gives the propensity that the input segment is related to a transmembrane region. A high output value (greater than 0.9) is used to trigger off the detection of a transmembrane segment. When at least one transmembrane region is detected, a protein is classified in the membrane protein class, otherwise it is put in the non-membrane protein class.

Experimentation determined the optimal size of the input layer to be 30. Training proteins were converted to input vectors by shifting a window of 30 residues successively through the sequence, i.e. the first segment contains amino acids from position 1 to position 30, the *nth* segment encodes amino acids from *n* to *n+30*. The training set was accordingly converted to 3140 input vectors of 30 values each.

Considering an input vector of 30 amino acids, we decide that it represents a significant transmembrane region if at least 10 amino acids in it belong to a transmembrane segment. Experimentation determined the optimal number of the hidden layers to be 1 and the number of neurons in this layer to be 2. The network was totally connected between adjacent layers.

The neural network was trained with the 3140 input vectors and their corresponding output values until convergence to a total error of less than 0.005.

**Results**

By using only information contained in 11 sequences from the training set, the neural network was able to generalize the processing to the testing set with a very good reliability. Applied on the 5 testing sets of membrane proteins (see methods), the system gives a perfect prediction rating of 100% by classifying all the sequences in the membrane class. For the test set of 995 globular proteins, the neural network predicts falsely 23 of them to be in the membrane class (97.7% of good assignment). The proteins falsely classified are: 1AGN, 1AMU, 1ARZ, 1AW8, 1BFD, 1BIB, 1BNK, 1CD1, 1DLC, 1FGJ, 1IHP, 1KVE, 1LXT, 1MAZ, 1NOX, 1OVA, 1PS1, 1TAD, 1TAH, 1UAE, 1WER, 2ABK and 3R1R.

These results are good but cannot be easily generalized to decide the predictive power of the method applied on real cases like the classification of Open Reading Frames (ORF) identified in complete genomes. The test sets of membrane proteins seems indeed too limited and composed exclusively of proteins with experimental and reliable information about the location of transmembrane segments. This doesn't necessarily reflects the composition of a complete genome. In addition, the proteins used in these sets contains only α-helices transmembrane which are easier to predict than β-strands transmembrane segments (Diederichs et al., 1998).

Despite the errors contained in SWISS-PROT, it is thought that the annotations contained in this database can be used to automatically extract two sets of membrane and non-membrane proteins which should be representative of the composition of complete genomes. These sets can serve as a common test that could be used for the rating and comparison of similar methods.

The set of membrane proteins extracted from the SWISS-PROT database release 37 contains 10743 entries. It has been built by selecting all sequences containing the keyword 'TRANSMEMBRANE' and having at least one transmembrane segment annotated. All remaining sequences in the database are not necessarily non-membrane proteins as some membrane sequences might be not yet annotated. A reliable set was extracted by selecting, from the cluster of proteins not marked transmembrane, only sequences with a known three-dimensional structure (presence of the '3D-STRUCTURE' keyword). The set of non-membrane proteins contains 2280 sequences.

The neural network was applied to these sets of membrane and globular proteins collected. For the membrane proteins, it correctly classifies 92.28% of them in the membrane protein set (9914 out of 10743). For the soluble proteins, it correctly classifies 93.38% of them (2129 out of 2280). The score obtained on this set of soluble proteins is lower than the rating calculated on the set listed in PDBSELECT but should be a good indication of the validity of the extraction method. The ratio of 93% of good assignment (both for membrane and non-membrane proteins) should be representative of the predictive power of the method when applied on complete genomes.

On the basis of these encouraging results, the neural network was associated with the PRED-TMR algorithm in a new application package called PRED-TMR2. This program can directly be used for the prediction of unknown proteins or on the ORF's (Open Reading Frames) predicted by the various genome projects.

**Table II.** Percentages of transmembrane proteins predicted by PRED-TMR2 on seven complete genomes.

| Genome names | % of TM proteins |
|---|---|
| Escherichia coli | 24.6 % |
| Haemophilus influenzae | 21.2 % |
| Methanococcus jannaschii | 19.8 % |
| Mycoplasma genitalium | 26.3 % |
| Mycoplasma preumoniae | 22.9 % |
| Saccharomyces cerevisiae | 28.0 % |
| Synechocystis SP | 26.5 % |

PRED-TMR2 has been applied on seven complete genomes and on the entire content of the SWISS-PROT database. The percentage of membrane sequences predicted in each genome is given in table II. The results range from 19.8% for Methanococcus jannaschii to 28% for Saccharomyces cerevisiae. Details of the results obtained can be downloaded together with the list of the transmembrane segment assignments from http://o2.db.uoa.gr/PRED-TMR2/Results/.

## Discussion

The prediction of transmembrane segments in proteins is a central domain in computational biology. A number of methods have been developed over the past 20 years. Some of them obtains good accuracy (Rost et al., 1995) and are available for biologist through the Internet. However, most of these methods are focused on the localization of α-helices transmembranes in known integral membrane proteins and produce a number of false segment detection when applied on soluble proteins. The rate of over-prediction is not well known as few publication were published on this subject. Two recent paper tackle the problem of identification of transmembrane proteins. Kihara et al. (1998) have tested their method on two sets of 89 transmembrane proteins collected in the literature and 928 globular proteins extracted from PDBSELECT. They announce a correct classification of 82 of the transmembrane proteins (92.13%) and of 836 of the globular ones (90.1%). Our neural network was found to perform slightly better than this methods. Hirokawa et al. (1998) made the tests of their SOSUI system on a set of 92 transmembrane proteins listed by Fariselli and Casadio (1996) and 502 soluble proteins extracted from PDBSELECT and state that their system discriminated all sequences correctly, except for one in each set of data ; resulting in an accuracy of more than 99%. Concerning the classification of transmembrane proteins, our method performs in a similar way as SOSUI as 100% of good classification were obtained on the same set. For the globular proteins, SOSUI with an incredible accuracy of 99.8% is far better than our neural network. However, an execution of SOSUI on the 23 soluble proteins misclassified by our system results in an assignment as transmembrane for 3 of them (1BNK, 1CD1, 1KVE). Even with these errors, the rating is still excellent, and better of our method if we consider that all remaining sequences are correctly predicted by SOSUI.

The systems above are not using neural network system for the classification. We show here that a simple neural network systems can be applied to this kind of problem in a successful way. The novelty in our network topology is the little number of neurons and connections required. Most of the neural network system presented so far use the same local encoding for each amino acid in a sequence (Qian and Sejnowski, 1988; Reczko, 1993; Rost et al., 1995; Fariselli and Casadio, 1996; Aloy et al., 1997; Diederichs et al., 1998), i.e. each residue is represented by a vector of 20 or 21 values. The input layer of the networks using this encoding must be 20 times the size of the input segment. In the case of a window of 30 amino acids, this represents 600 neurons. In our system, each amino acid is encoded with an unique value and only 2 neurons in the hidden layer are used.

It is known that a successful generalization of a prediction by a neural network requires a much larger number of cases that the number of weights adjusted during the training phase. With our architecture, the total number of connections associated with a weight is only 62 (60 to connect 30 input neurons to the hidden layer and 2 from this layer to the unique output). This allows to successfully train the network with information on the topology of very few proteins. In our application, the number of cases (3140) is a factor of 50 higher than the number of weight.

In addition, the simple feed-forward topology of the network and its limited number of connections allow proteins to be processed very quickly and could open the way for new implementation able to handle longer segments of amino acids and, why not, complete sequences.

A WWW server running the PRED-TMR2 algorithm is available at http://o2.db.uoa.gr/PRED-TMR2/

## Acknowledgements

The authors gratefully acknowledge the support of the EEC-TMR "GENEQUIZ" grant ERBFMRXCT960019.

# References

Aloy, P., Cedano, J., Olivia, B., Aviles, X. and Querol, E. (1997) *CABIOS*, **13(3)**, 231-234

Bairoch, A. and Apweiler, R. (1998) *Nucleid Acids Res.*, **26**, 38-42

Cserzo, K., Wallin, E., Simon, I., von Heijne, G. and Elofsson, A. (1997) *Protein Engineering*, **10(6)**, 673-676

Diederichs, K., Freigang, J., Umhau. S., Zeth, K. and Breed, J. (1998) *Protein Science*, **7**, 2413-2420

Fariselli, P and Casadio, R. (1998) *Comput. Applic. Biosci.*, **12(1)**, 41-48.

Hirokawa, T., Boon-Chieng, S. and Mitaku S. (1998) *BioInformatics*, **14(4)**, 378-379

Hobohm, U. and Sander, C. (1994) *Protein Science*, **3**, 522

Kihara, D., Shimizu, T. and Kanehisa, M. (1998) *Protein Engineering*, **11(11)**, 961-970

Pasquier, C., Promponas, V.J., Palaios, G.A., Hamodrakas, J.S. and Hamodrakas, S.J. submitted for publication

Persson, B. and Argos, P. (1994) *J. Mol. Biol.* , **237**, 182-192

Qian, N and Sejnowski, TJ. (1988) *J. Mol. Biol.*, **262**, 865-884

Reczko, M. (1993) *SAR and QSAR in Environmental Research*, **1**, 153-159

Rost, B., Sander, C. and Schneider, R. (1994) *CABIOS*, **10(1)**, 53-60

Rost, B., Casadio, R. Fariselli, P. and Sander, C. (1995) *Prot. Sci.*, **4**, 521-533

Rost, B., Fariselli, P. and Casadio, R. (1996) *Prot. Sci.*, **5**, 1704-1718

Sussman, J.L., Ling, D., Jiang, J., Manning, N.O., Prilusky, J., Ritter, O. and Abola, E.E. (1998) *Acta Cryst.* **54**, 1078-1084

Von Heijne, G. (1992) *J. Mol. Biol.*, **225**, 487-494